\documentclass[runningheads]{cl2emult}

\begin{document}
\title*{OPTICON and the Virtual Observatory}

\toctitle{OPTICON and the Virtual Observatory}

\titlerunning{OPTICON and the Virtual Observatory}

\author{Gerry Gilmore}

\authorrunning{Gerry Gilmore}

\institute{Institute of Astronomy, Madingley Rd, Cambridge CB3 0HA, UK
                      \\ gil@ast.cam.ac.uk; www.ast.cam.ac.uk/~gil/}

\maketitle              % typesets the title of the contribution

\begin{abstract}
The challenges of multi-wavelength astrophysics
require new outlooks from those appropriate to traditional astronomy.
The next generation of research scientists must be trained to exploit
the potentiality now being provided for the first time. Just as
importantly, the full range of available information must be indexed
and made available, to avoid wasteful repeat observations, or
incomplete analyses. Perhaps the greatest challenge in the immediate
future is to ensure the wealth of multi-wavelength data already
available, and being accumulated, is available for efficient
scientific exploitation. The difference between observations in a
depositary and a fully-operational data archive is the difference
between waste and cutting-edge science. The EU Optical Infrared
Coordination Network for Astronomy (OPTICON) provides a forum to
coordinate and develop the many national and international efforts and
desires leading towards an operational virtial observatory.
\end{abstract}

\section{Do We Need a Virtual Observatory?}

The title of this section introduced a panel discussion at the end of
the exciting, innovative and important meeting, Mining the Sky. The
remarkable feature of the discussion was the overwhelming agreement that
the case for an international Virtual Observatory is simply
unassailable, even though its implementation will be an enormous, and
expensive, challenge. In spite of that confidence, much effort is
required first to produce a specific case for the virtual observatory
in order to fund it, and second to build and operate it.

The topic of the virtual observatory has been the subject of much
recent discussion, and need not be repeated here in general terms 
(for an updated www site see www.astro.caltech.edu/nvoconf/). Instead
I record a few of the challenges which were raised during the
discussion session, and describe below some of the European-level
activity which is underway today.

\subsection{Why?}

There was wide agreement that the most beneficial consequences of a
virtual observatory, judged from the perspective of the international
astrophysical community, will be
\begin{itemize}
\item New Science
\item More Science
\item Better Science
\item Less Waste.
\end{itemize}
It is much less obvious that these same benefits are those which will
support funding proposals.

\paragraph{New science} will follow from bringing together more information in a
more consistent way. Astronomy works like that: people discover new
things, and new classes of object, every time new facilities are
allowed by technology. We share a discovery-led subject, not a
theory-led one. This serendipity argument is however hard to quantify,
and therein lies its weakness for funding proposals.
One funds a new telescope or computer largely on some predicted
threshold which can be crossed; with hindsight the most interesting and
important discoveries are not those in the original science
case. Perhaps the most difficult immediate challenge for the virtual
observatory is to present the science case we all know is strong in
terms which will appeal to funding agencies.

\paragraph{ More science} will follow because the user community able
to conduct state of the art research will be vastly increased. This,
an aspect I personally consider the most important, is particularly
important in an expanding European Union. There are substantial
numbers of extremely talented and well educated scientists in, for example, the
countries of central Europe and the FSU, who lack access to expensive
technology. In the internet age every university could have access to
state of the art facilities, for merely the communication cost. This
argument however naturally appeals more to international than to
national funding agencies.

\paragraph {Better Science} will follow the combination of the two
points above, and also by the development of the new tools essential
for data mining and virtual facilities. While our computational
facilities are growing as Moore's Law, my brain is not. Without the
focussed effort required to meet the virtual observatory challenge,
especially in development of analysis and access tools which are
transparent to the user,
we will waste the information potentially available, but un-mined, in our data.

\paragraph{ Less waste} is an inevitable, and happy, consequence of
the maximally effective utilisation of investment. The more data are
independently used, analysed, interpreted, the more value is attached
to each observation. So long, of course, as the re-analyses are
cheaper than is acquisition of new data. The virtual observatory, like
everything else on this scale, is going to require dedicated staff,
career structures, and lead organisations. A continuing challenge will
be to ensure that the organisation remains at the leading
edge of technology. The existing successful data centres show that
this challenge can indeed be met, but not at minimal recurrent cost.

Three other points were raised in the discussion,
\begin{itemize}
\item What?
\item Who pays?
\item How?
\end{itemize}

\paragraph{ What} should be in the virtual observatory? The answer
here is a challenge still to be met! There is a good case that popular
simulations should be readily available, with appropriate analysis
tools. Given the rate of change of such things, however, deciding what
to support without the benefit of hindsight remains problematic. The
same situation occurs with data: there is a distinction between a
large archive and a good archive. Huge amounts of badly calibrated
data are of little use, while small high-quality data may be extremely
desirable. One may compare the costs of access to the
works of Shakespeare (small, cheap, rich in interpretive value) and
the Iridium satellite system (fancy access possibilities, less obvious
demand). There is a general concensus that the virtual observatory
will start with well-calibrated surveys and dedicated complementary
information, but a collection of surveys is no more a virtual
observatory than a mirror is a telescope.

\paragraph{ Who pays?} Who decides? Funding international access is
not easy in some countries, but is a natural requirement for some
agencies, such as the EU. The existence of international
organisations, benefactors with humanitarian interests, and the
internet shows these challenges, while hard, can be met.

\paragraph{How} does one make the virtual observatory?  We are
fortunate here in that the lead has been taken by the bigger subjects,
such technical developments as the GRID, and development of
technological solutions on the appropriate scale (eg www.globus.com).
It is also reassuring that in astronomy we have available some
excellent role models. Such facilities as the  HST archives/ECF, CDS,
and such like are an important and excellent first step on which we
can develop. The European activity noted below utilises a
step-development approach, building on experience, and driven by
science return.

\section{OPTICON, the EU Optical Infrared Coordination Network for Astronomy}

Many of the European-level activities working towards the virtual
observatory are sponsored or coordinated by the EU OPTICON
Coordination Network. Coordination networks, of which OPTICON is one,
are supported by the EU in each of the many sub-fields in which the EU
funds research. The primary goal of such networks is to provide a
forum for the national agencies, together with some users, to develop
and coordinate joint approaches to common challenges. In addition to
improved coordination of research and planning at a European level,
which is one of its goals, a further benefit for
the EU is provision of information from the national communities back
to the EU, 
allowing improved definition and implementation of future EU
programs. The general projects are described on the EU WWW site \\
www.cordis.lu/improving/src/ari$_{-}$th.htm.

The infrastructure for optical and infrared astronomy in Europe is
operated by a combination of multi-national organisations and a number
of national and regional funding agencies. This infrastructure is in a
time of rapid technological development, raising new challenges common
to all facilities and users.  

OPTICON, the coordination network in European optical-infrared
astronomy, brings together the major European national and
multi-national operators of optical astronomical observatories and
archives, and several major research groups containing users of
infrastructure.  Through OPTICON, the national agencies are addressing
three common goals: to provide cost-effective use of the substantial
financial investment in telescopes and archives for the widest possible
user-community; to maximise the operational efficiency of extant small
and medium sized telescopes; and to develop proposals for future very
large facilities. The virtual observatory is a key part in achieving
these goals.

\paragraph{OPTICON PARTNERS, and Contact Names}

\begin{itemize}
\item{}1) PPARC (UK; Administrative coordinator) (Paul Murdin, Ian Corbett)
\item{}2) IoA Cambridge (Scientific coordinator; Chair) (Gerry Gilmore)
\item{}3) CDS (Strasbourg, France)(Francoise Genova, Daniel Egret)
\item{}4) ESA Space Sciences Division (including ST-ECF) (Piero Benvenuti)
\item{}5) ESO (Alvio Renzini)
\item{}6) IAParis (Bernard Fort, Alain Omont)
\item{}7) CNRS/INSU (France) (Genevieve deBouzy)
\item{}8) Instituto de Astrofisica de Canarias (Francisco Sanchez, Rafael Rebolo)
\item{}9) Italian National Consortium for Astronomy and Astrophysics
(CNAA) (Marcello Rodono, Giancarlo Setti, Franco Pacini)
\item{}10) Leiden University (George Miley, Harm Habing)
\item{}11) MPfA (Garching), (Simon White)
\item{}12) MPiA (Heidelberg; and Calar Alto Obs) (Hans-Walter Rix)
\item{}13) NOVA  Nederlandse Onderzoekschool Voor Astronomie (Tim deZeeuw)
\item{}14) Nordic Optical Telescope Scientific Association (Vilppu
Piirola, Leo Takalo)
\end{itemize}

Efficient and effective scientific exploitation of the combination of
new large telescopes and new wavelengths is presenting entirely new
challenges to the European astronomical community.  At an operational
level, the cost of these facilities strains national research budgets,
requiring efficient planning and multi-national coordination in their
present operation and future instrumentation. The continuing efficient
use of mid-sized facilities requires consideration of their
complementarity to larger, more specialised, facilities. Future new
developments are necessarily multi-national, requiring efficient
community interactions at preliminary planning levels. Consideration
must include ways in which countries without major national
investments can be integrated into the astronomical community, especially
in view of future EU enlargement. All these plans
require close interaction between the infrastructure operators, and
experienced researchers, to ensure operational modes are jointly
enhanced for the benefit of the research community overall.

OPTICON sponsored workshops are developing the science case for the
next generation Extremely Large telescopes
(www.roe.ac.uk/atc/elt/) for coordinated operation of Europe's 2-4m
telescopes, and for a variety of common infrastructure and
interoperability issues. Considerable activity is underway relevant to
the Virtual Observatory, a major priority for OPTICON.

Because astronomy is fundamentally an observational, rather than
experimental, science, an archived observation is frequently as
valuable as is a new one. Additionally, multi-wavelength archived
information is frequently essential to full astrophysical analysis of
a specific source.  Delivery of this capability implies that
appropriate multi-wavelength data be maintained and supported in an
archive, be indexed and cross-linked, and be readily accessible, in a
standard format, to the user community with realistic effort. In
general, Europe has an outstanding record in provision of such
facilities, with the Centre de Donnees astronomiques de Strasbourg
(CDS), the Garching Space Telescope European Coordinating Facility,
ESO, and several other large archives being of leading international
stature. Many collaborations and cooperative developments inside
Europe and with the US are underway, while huge new developments are
planned for new large telescope and space mission archives. All these
need to be coordinated with each other, with the many extant
observatory archives, with the forthcoming large survey archives, and made
compatible with rapidly changing communications and computing
technologies. Optimised communications between archive developers,
observatories, and users is critical, and mutually beneficial for
operators and for users.

The approach adopted is very much bottom-up, incremental, and results
led.  The immediate  goals are to
quantify the science requirements on the basis of actual research
plans, and to implement true inter-operability between extant major
data archives. Only after real new science has been delivered by
accessing real archives can realistic extensions to a true virtual
observatory be defined, costed and implemented. 

The first part of this is being implemented through an EU-funded
program, ASTROVIRTEL, led by P Benvenuti
(Piero.Benvenuti@eso.org). The second builds on the very considerable
experience developed at CDS Strasbourg (cdsweb.u-strasbg.fr) over many
years in delivering astronomical data and publications on-line to an
open user community. A small example of the complexities of real-world
interoperability challenges which must be solved by a virtual
observatory is access to tabular data. The documents at
vizier.u-strasbg.fr/doc/catstd-1.htx make salutary reading, while a
special issue of Astronomy and Astrophysics, volume 143 number 1, pp
1-143, April 2000, and which also describes the NASA ADS facility, is
available for those who prefer paper, or to read on airplanes.

Opticon supported working groups, involving the international
community, and so ensuring close collaboration with North American
activities, are being led by CDS Strasbourg
(genova@astro.u-strasbg.fr), to develop suitable standards and
interfaces for future improved performances.

Future progress towards the virtual observatory will be slow, and
require a lot of work and resources, much of which may be provided by
the EU. In this respect astrophysics is following the lead of particle
physics, earth observation, and biology, which have already EU-led
proposals involving their equivalent of OPTICON-like collaborations
(see for example grid.web.cern.ch/grid/).

\subsection{ ASTROVIRTEL; ecf.hq.eso.org/astrovirtel/}

The ESO/ST-ECF Archive currently contains more than 7.0 Terabytes of
scientific data obtained with the ESA/NASA HST, with the ESO NTT, VLT
and with the Wide Field Imager on the ESO/MPI 2.2m Telescope. The
growth rate is 4.5 Tbytes per year ramping to 6.0 Tbytes/y within the
next two years. In addition to 'public' data arising from General
Observer programmes whose one-year proprietary period has elapsed, the
HST and ESO Archives contain some large datasets resulting from
programmes approved with a reduced or nonexistent proprietary period -
this includes 'parallel' data from a second instrument obtained
simultaneously with pointed observations by the primary instrument,
the ESO Imaging Survey and VLT Science Verification and Commissioning
data. Large public datasets are expected to accumulate ever faster
during the second decade of HST operations and after operations start
for ground-based, wide-field images like the VLT Survey Telescope.

In addition, network connection allows Archive Users to retrieve data
from other active Archive Centres (e.g. ISO). Intermediate data sets
can be staged on fast access robotic devices. This unique (in size and
quality) collection of diversified astronomical data can now be seen
as a "virtual" observatory, capable of responding to requirements for
observations as a "real" first-class telescope.

The ASTROVIRTEL Project, supported by the European Commission and
managed by the ST-ECF on behalf of ESA and ESO, is aimed at enhancing
the scientific return of the ESO/ST-ECF Archive. It offers the
possibility to European Users to exploit it as a virtual telescope,
retrieving and analysing large quantities of data with the assistance
of the Archive operators and personnel.

A Selection Panel will select up to six Proposals per year. The
selection criteria will be:
\begin{itemize}
\item       Scientific originality and excellence 
\item       Scientific exploitation of large and multi-instrument data sets 

\item       Particular attention will be given to those research programmes
requiring the development of new "mining tools" that might be of
general interest to the community.
\end{itemize}

For the selected proposals ASTROVIRTEL will provide:
\begin{itemize}
\item        Travel and subsistence for PI and CoI visits to the ESO/ST-ECF
       Archive Facility for discussing the proposal and planning the
       development and operation of relevant tools.  
\item Support in the
       development and operation of the "mining tools", specifically
       designed for the proposal 
\item Support in the retrieval of
       additional data sets, when not locally available 
\item Support for the analysis and interpretation of the data
\end{itemize}

Comments and examples.

Up to now, HST Archival Proposals were the prerogative of US
Astronomers who can submit them within the normal HST Observing
Cycles. The ASTROVIRTEL Project is partly aiming at filling this gap,
with two notable differences:
\begin{itemize}
  \item      The Archival Proposals are not limited to HST data. 
\item        The support is mainly offered as services to the
Proposers, rather than as pure grants. 
\end{itemize} 
The advantages of the ASTROVIRTEL approach are that: 
\begin{itemize}
\item        the "scientific interoperability" of different Archives will be
       enhanced on the basis of specific scientific requirements
       (those contained in the approved Proposals); 
\item the "mining tools"
       and the procedures for the management and analysis of the
       retrieved data sets will become part of the Archive and offered
       to the community.
\end{itemize} 

An example may clarify the concept: let's assume that the scientific
aim of an approved Proposal is to derive photometric redshifts from
suitable deep HST images. The PI and CoIs (typically two scientists in
total) will be invited to spend a few days with the Science Archive
support group in order to define the best strategy for the selection,
retrieval and pre-processing of the data (travel and subsistence will
be covered by ASTROVIRTEL). The visit will be preceded by extensive
electronic communication so that preliminary testing of the different
procedures can be run and discussed during the visit (e.g.  selection
of fields sufficiently deep and with suitable photometric coverage,
"drizzling" and cosmic ray removal, tuning the parameters for
extraction of objects, correlation/identification of extracted objects
in different colour frames, estimate of colours and errors,
compilation of object catalogues and correlation with other
observations and/or catalogues, etc. etc.). After this first visit and
on the basis of the tests and of the discussion, the Archive Support
Group will support the joint development and implementation of a
working version of the specific tools and procedures and help to test
them on the Archive(s). Intermediate steps will be discussed
electronically with the PI Group. A second visit (also supported by
ASTROVIRTEL) of the PI Group should be envisaged toward the conclusion
of the retrieval and the pre-processing of the data. The Archive Group
will later make the tools and procedures more robust, document them
and offer them to the generic Archive Users.

\section{Conclusion}

Archives contain a mixture of data types (images, spectra, tables), of
data states (raw data, reduced and calibrated spectra, flux tables,
...), of formats, of means of access, of ease of cross access
(multi-wavelength data on a single object is stored in several
different archives, in different formats, and different states of
calibration), and different types of indexing (minimal positional
information to project-specific naming conventions).  Coordination
between the several existing and forthcoming data archives is clearly
of direct benefit to overall productivity and access, as are programs
to enhance the practical access to archives.

To get the maximum and best science, the full range of
facilities must be made available to the best scientists, with
adequate tools to ensure their efficient use.  While much of science
proceeds through healthy competition, some developments, particularly
of common infrastructures, are naturally coordinated across countries
and communities. 

The participants at the Mining the Sky meeting were clear that a
virtual observatory is inevitable, and desirable. Partly that is a
consequence of the selection effect of the meeting subject. But only
partly: the scientific, financial, and demographic cases for
development of a virtual observatory are strong, and very widely
recognised. Implementation of a viable virtual observatory will however
be expensive, and more importantly will be technically
challenging. Implementing a huge state of the art technological
capability will strain the communities resources. My personal belief is
that the virtual observatory will grow by bootstrap techniques: as
those communites which have least current infrastructure (eg central
Europe) and so which will be the greatest relative 
beneficiaries, begin access, they will fine tune the tools, implement
the user-friendly aspects, and make the project real.

It is the task of extant multinational forums, such as the OPTICON
coordination network, to ensure that a basic set of conventions is
developed and implemented, so that the virtual observatory will
deliver its exciting potential: allowing the whole community to mine
the sky.

%INDEX%%%%%%%%%%%%%%%%%%%%%%%%%%%%%%%%%%%%%%%%%%%%%%%%%%%%%%%%%%%%%%%
%\clearpage
%\addcontentsline{toc}{section}{Index}
%\flushbottom
%\printindex
%%%%%%%%%%%%%%%%%%%%%%%%%%%%%%%%%%%%%%%%%%%%%%%%%%%%%%%%%%%%%%%%%%%%%

\end{document}